\documentclass[a4paper,twoside]{article}

\usepackage{epsfig}
\usepackage{subcaption}
\usepackage{calc}
\usepackage{amssymb}
\usepackage{amstext}
\usepackage{amsmath}
\usepackage{amsthm}
\usepackage{booktabs}
\usepackage{multicol}
\usepackage{pslatex}
\usepackage{apalike}
\usepackage[bottom]{footmisc}
\usepackage{listings}
\usepackage{url}

\def\rot{\rotatebox}

\usepackage{comment}
\usepackage[outdir=./]{epstopdf}

\usepackage{SCITEPRESS}     
\usepackage{tikz}

\newcommand{\cmark}{%
\tikz[scale=0.23] {
    \draw[line width=0.7,line cap=round] (0.25,0) to [bend left=10] (1,1);
    \draw[line width=0.8,line cap=round] (0,0.35) to [bend right=1] (0.23,0);
}}

\begin{document}

\title{SCANTRAP: Protecting Content Management Systems from Vulnerability Scanners with Cyber Deception and Obfuscation}

\author{\authorname{Daniel Reti\sup{1}\orcidAuthor{0000-0001-8071-6188}, Karina Elzer\sup{1}\orcidAuthor{0000-0002-3984-779X} and Hans Dieter Schotten\sup{1,2}\orcidAuthor{0000-0001-5005-3635}}
\affiliation{\sup{1}German Research Center for Artificial Intelligence, Kaiserslautern, Germany}
\affiliation{\sup{2}Technische Universität Kaiserslautern, Germany}
\email{\{firstname\}.\{lastname\}@dfki.de}
}


\abstract{Every attack begins with gathering information about the target. The entry point for network breaches are often vulnerabilities in internet facing websites, which often rely on an off-the-shelf Content Management System (CMS). Bot networks and human attackers alike rely on automated scanners to gather information about the CMS software installed and potential vulnerabilities. To increase the security of websites using a CMS, it is desirable to make the use of CMS scanners less reliable.
The aim of this work is to extend the current knowledge about cyber deception in regard to CMS. To demonstrate this, a WordPress Plugin called 'SCANTRAP' was created, which uses simulation and dissimulation in regards to plugins, themes, versions, and users.
We found that the resulting plugin is capable of obfuscating real information and to a certain extent inject false information to the output of one of the most popular WordPress scanners, WPScan, without limiting the legitimate functionality of the WordPress installation.}

\onecolumn \maketitle \normalsize \setcounter{footnote}{0} \vfill

\section{\uppercase{Introduction}}
\label{sec:introduction}
    Content Management Systems (CMS) have become a notable part of today's web, with almost half of the websites using some sort of CMS. This is due to the ease of use and free option. Specifically, out of the existing web CMSs, some popular CMSs have emerged: WordPress (WP) is the most popular CMS, with a market share of 63.3\%. Shopify, Wix and Squarespace, as paid alternatives, together have 12.4\% of the market share. The most popular open-source alternatives, Joomla and Drupal, are used in 4.3\% of websites using CMS. \cite{CMSStats}
    
    With such popularity, however, they also become an attractive target for attackers. Due to the addition of third-party extensions, these platforms leave a large attack surface. In the field of penetration testing, different tools have emerged that reduce the effort of scanning and exploitation of popular CMSs. This makes reconnaissance possible with low effort, enabling the threat from so-called 'script kiddies' and bots.
    
    Therefore, it is desirable to increase the effort of attackers in regards to attacking CMS by using different defence strategies. This starts with the attacker's capability of scanning CMSs. The approach of this work is based on cyber deception to disguise valuable information, by hiding the real and presenting false information.

    As the most popular CMS as of today, we choose WP as a representative CMS to apply deception measures for our research. Due to its popularity, almost 100,000 plugins and 25,000 themes exist for WP. This leaves a large attack surface leading to over 30,000 known vulnerabilities. 92\% of which are found in plugins. \cite{WPVulnstats} Also, because of its popularity, we choose WPScan, a WP vulnerability scanner maintained by Automattic, the vendor of wordpress.com, as the attacker's tool of choice.

    This paper presents a systematic approach to deception based defence against CMS scanners. Hereby a WP plugin to defend against WPScan is created as a proof of concept. This plugin, called 'SCANTRAP', uses simulation and dissimulation techniques of the cyber deception domain to make reconnaissance harder for attackers. The source code can be found at https://github.com/dfki-in-sec/SCANTRAP.
    
    Additionally, we give an overview of cyber deception methods for the web domain, the functionality of different CMS scanners and a survey of cyber deception methods used by existing WP security plugins.
    
    Our paper is divided into six sections. The following section gives a brief overview of the background of cyber deception and CMS. The third section describes our created plugin. In the fourth, we present our discussion and in section five we examine related work. Our conclusion is drawn in the final section.

\section{Content Management System Security and Cyber Deception}
\label{sec:background}
    In the cyber kill chain, a model describing the stages of an attack, the first phase is reconnaissance. In this phase, the attacker will gather information that will allow clearer decision-making in later steps. One method of reconnaissance is scanning. This paper's focus lies on defending this part of the cyber kill chain through cyber deception in the domain of CMS, which will be explained in detail in this section.

    \subsection{Cyber Deception}
    \label{sec:cyber_deception}
        In the field of cyber security, the term \textit{cyber deception} is generally understood as a defence strategy which is used to intentionally mislead an attacker. Specifically, by simulating fake or hiding existing targets, the actions and in-actions of the attacker should become more predictable or take more time. \cite{cyber_deception}
        
        In the literature, many deception techniques in mulitple computer domains have been proposed and studied \cite{fraunholz.2018}. 
        
         In 1991, Bell and Whaley defined a taxonomy for deception that contains two main classes \cite{deception_tax}. Dissimulation describes the hiding of the existing and simulation refers to faking.
        
        The term 'obfuscation' is a type of cyber deception and dissimulation, mainly understood in relation to code protection. However, the idea is to transform a structure to make it harder for an attacker to determine the functionality. \cite{BEHERA2015757}
        
        More concrete examples of cyber deception are honeypots and honeytokens. Honeypots are fake computer systems that allow the detection of unusual interactions. Honeytokens serve the same purpose but are anything but a computer. \cite{de_barros_2007} \cite{de_barros_2007} \cite{spitzner_2003}
        
        Regarding cyber deception for web services, multiple specific techniques can be implemented. In this work, the following methods were considered, based on the work of \cite{cloxy}:
        \begin{itemize}
            \item \textit{Version Trickery} is a form of dissimulation that provides falsified version information. This could be the version of the HTTP Server, the Operating System, the CMS installation, plugins or themes. This makes it harder for the attacker to identify vulnerable versions or find exploits. In the context of CMSs, it is possible to obfuscate the version completely, as described in Section \ref{sec:version_detection}. For WP, exisiting plugins can fake the version, see Table \ref{tab:secplugins}.
            
            \item \textit{Disallow Injection} refers to injecting \textit{Disallow} entries in \textit{robots exclusion file}. These entries are used by web crawlers as well as different scanners, such as CMS Scanners, to find possibly sensitive paths and files. This falls into the category of simulation and can be used for detection purposes. The WP plugin 'Blackhole for Bad Bots' \cite{blackhole} is an example for faking entries in the \textit{robots} file.
            
            \item \textit{Status Code Tampering} refers to manipulating the HTTP response codes and could be used agaist scanners relying on this status code to test whether a certain path exists or not. It was used in this work to hide and fake the existence of specific plugins and is further described in Section \ref{sec:plugin_detection}.
            
            \item \textit{Latency Adaption} is a simulation tactic to prevent brute-force attacks and can be thought of as a veiled rate limiting. The more requests are sent, the more the response from the server will be delayed, which limits the number of requests that the attacker can send. Rate-limiting brute force login attacks is a method used by different CMS security plugins (Sec. \ref{sec:related_work}).
            
            \item \textit{Virtual Honey Files} is a method where invented resources on the website are presented, with the goal of creating uncertainty for the attacker. This could be a fake directory, document, login page, back end, CMS installation etc. In this work a similar method was used to include plugin version numbers as described in  \ref{sec:plugin_detection}.  
            
            \item \textit{Code Obfuscation} is used to hide the functionality of code such as  \textit{JavaScript}. This is used to prevent the stealing of intellectual property. It might be effective against CMS scanners when specific patterns in the HTML are analysed to extract information, but has not been applied in this work.
            
            \item \textit{Cookie Scrambling} can be used as either simulation or dissimulation tactic. This is done by either creating a honey cookie, where tampering with the cookie can be detected or by creating cookies that need to be set to successfully execute a task. An example of this for Wp is described by a blog post \cite{medium_talk}, which is explained more closely in Section \ref{sec:related_work}. 
            
            \item \textit{Content Modification} can be used for dissimulation and simulation. Example for content modification are described in Sections \ref{sec:version_detection} and \ref{sec:user_enum} where users and versions are hidden. Furthermore, as seen in Table \ref{tab:secplugins}, it can also be used to fake users and the version.
        \end{itemize}
        
    \subsection{CMS}
        A Content Management System (CMS) is a way of managing content, which includes the collection and usage of the content. This allows for consistency and less repetition for the user. \cite[p. 72]{cm_bible} There are two types of CMS: enterprise CMS, which contains content relevant to the whole organisation, and web CMS's, which focus is on managing content for websites. \cite[p. 79-82]{cm_bible}
        
    \subsection{CMS Scanner}
    
        \begin{small}
        \begin{table*}[t]\centering
        \caption{Comparing different CMS scanners functionality \cite{CMSScanner_comp}.}
        \label{tab:cmsscanner} 
        {\def\arraystretch{1}\tabcolsep=0.6pt
            \begin{tabular}{|c|c|c|c|c|c|c|c|c|c|c|} 
             \hline
             Scanner & CMS & Module & Version & User & Misconfigs & URLs & Exposed & Login & Exploits & Code  \\
             & & Enumeration & Detection & Enumeration & & & Files & Attacks & & Analysis \\
             \hline\hline
             Droopescan & multiple & \checkmark & \checkmark & - & - & \checkmark & \checkmark & - & - & - \\ 
             \hline
             CMSmap & multiple & \checkmark & \checkmark & \checkmark & \checkmark & \checkmark & \checkmark & \checkmark & - & - \\ 
             \hline
             CMSeek & multiple & \checkmark & \checkmark & \checkmark & \checkmark & - & \checkmark & - & - & - \\ 
             \hline
             WPScan & WordPress & \checkmark & \checkmark & \checkmark & \checkmark & \checkmark & \checkmark & \checkmark & - & - \\ 
             \hline
             WPSeku & WordPress & \checkmark & - & - & \checkmark & - & \checkmark & \checkmark & - & \checkmark \\
             \hline
             JoomScan & Joomla & \checkmark & \checkmark & - & \checkmark & - & \checkmark & - & - & - \\ 
             \hline
             JoomlaVS & Joomla & \checkmark & \checkmark & - & \checkmark & - & - & - & - & - \\ 
             \hline
             JScanner & Joomla & - & \checkmark & \checkmark & - & - & - & - & - & - \\ 
             \hline
             Drupwn & Drupal & \checkmark & \checkmark & \checkmark & - & - & \checkmark & - & \checkmark & - \\ 
             \hline
            \end{tabular}
        }
        \end{table*}
        \end{small}
    
        CMS scanners are a type of vulnerability scanner. Mainly, they enumerate a CMS to find as much information that can be used to determine vulnerabilities. There are different open-source scanners. Some scanners work for multiple different CMSs others are for a specific CMS.
        
        In Table \ref{tab:cmsscanner}, a comparison was made between different CMS scanners and their functionalities. Module enumeration depends on the CMS and is generally done through directory brute forcing. For example, in WordPress, it would be plugin and theme enumeration. With one exception, every scanner has this functionality since these modules prone to contain vulnerabilities. Version detection shows similar results. User enumeration is possible for five out of the nine scanners. The detection of misconfigurations can be done by six scanners. URLs refer to the detection of 'interesting' URLs that are not necessarily searched for. Exposed files include backup, default, log and other files. Three scanners can do brute-force login attacks. Drupwn is the only scanner that has the integrated possibility of also exploiting found vulnerabilities, while some of the other scanners might link to possible vulnerabilities. Finally, WPSeku allows for static code analysis, which no other scanner offers. \cite{CMSScanner_comp}
    
    \subsection{WPScan}
        WordPress is the most popular, free and open-source CMS on the market. It allows the creation of websites and blogs with high customizability through a wide range of community-created themes and plugins, which allow for customized style and functionality, respectively.

        WPScan is one of the most known and advanced CMS scanners for WordPress with a large feature set (Tab. \ref{tab:cmsscanner}). In the work the focus is on its capabilities of theme, plugin, user and version enumeration and detection.

\section{SCANTRAP WPScan Evader}
    For evaluating the ability to evade CMS scanners, we used WordPress as the CMS and WPScan as the CMS scanner and created a plugin for WordPress that focused on the four main scans of WPScan. A the source code of WPScan is publicly available \cite{WPScanGithub}, it was used to better understand how the scans are performed and how the information is gathered.

    Although WPScan reveals itself by setting the HTTP User-Agent field to 'WPScan', this information was not used, as it could be easily changed for malicious purposes. Instead, cyber deception was used, as explained in \ref{sec:cyber_deception}, to manipulate the scanner output.
    Additionally, we partly adapted and modified ideas from a blog post \cite{WPScanEvasion} on this matter.

    \subsection{Plugin Detection}
    \label{sec:plugin_detection}
        Plugin detection is how WPScan finds existing plugins as well as their versions. The focal method for finding plugins is based on known locations. The WordPress structure has a determined folder ('/wp-content/plugins/') for plugins. Therefore WPScan does a directory scan for known plugins in this folder. If the directory for the plugin exists, which is based on the response code, then the mandatory 'readme.txt' file is requested and the version is parsed from either the 'Stable Tag' or 'ChangeLog Section'.
    
        For this work, two methods to evade plugin detection were used. The first approach is based on simulation. We faked plugins by changing the 404 response code of requests to non-existing plugins. WPScan determines a plugin as existing, when one out of four response codes is detected. The code "200 OK" means the request succeeded. "401 Unauthorized" stands for unauthenticated access and "403 Forbidden" means the user has no access rights. Lastly, "500 Internal Server Error" indicates an error on the server's side. For responses to plugin folder requests, we chose the 403 response code and for the 'readme.txt' file we selected status code 200. Further, we added customised text to the response body to add a stable tag for version detection. As an addition, based on honeytokens, we added simple logging when fake plugin paths are requested. This extends deception as a defence to also be able to detect unusual behaviour. An alternative to logging could have been an email notification.
        
        The second method of hiding the existing plugins, is a method of  dissimulation. The plugins are hidden by responding to requests to the plugin folder with a '404 Not Found' response. This is possible because WPScan will request files only when the folders are found. It still allows WordPress to use the contents of the folders, therefore not restricting the functionality of the plugins.
    
\begin{figure*}
            \centering
            \includegraphics[width=0.8\textwidth]{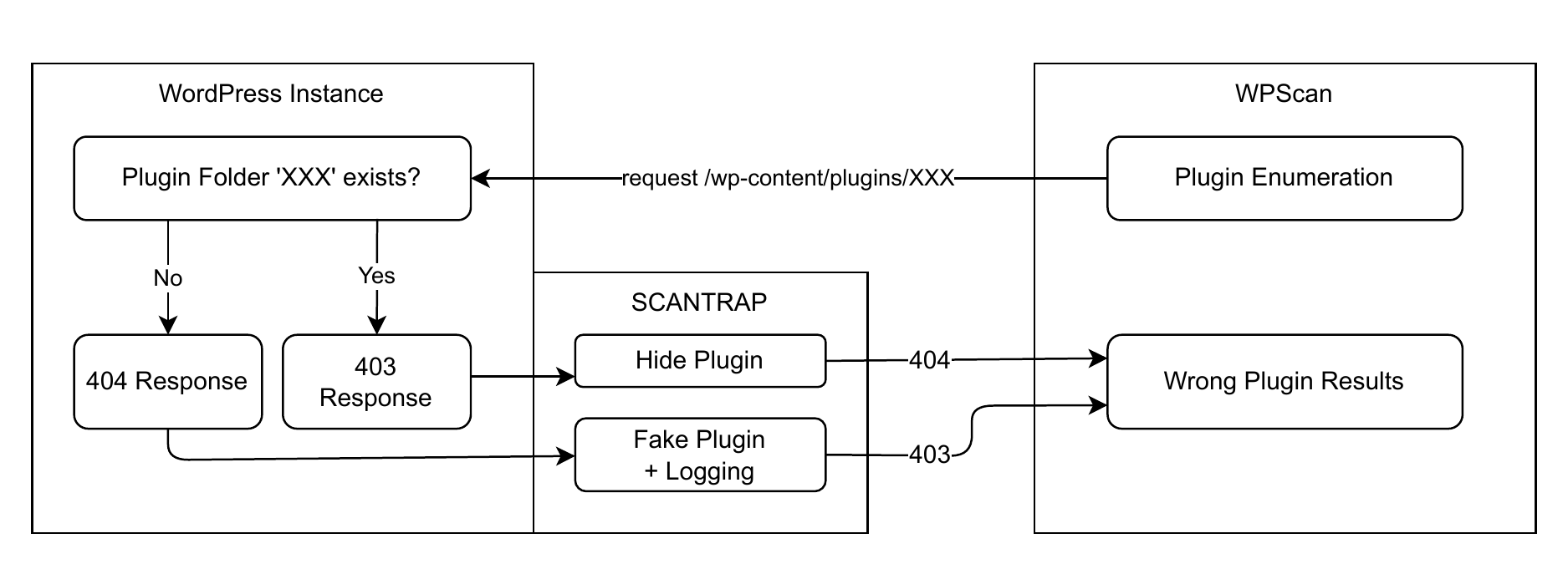}
            \caption{Demonstration of response manipulation of request for plugin folder 'XXX' with WP plugin 'SCANTRAP'.}
            \label{fig:plugin_enum}
        \end{figure*}

        Figure \ref{fig:plugin_enum} shows the sequence of request and response flow. It illustrates the manipulation of responses for plugin folder requests as described above.  
    
    \subsection{Theme Detection}
    \label{sec:theme_detection}
        Theme detection is identical to plugin detection, only with a change of the folder ('/wp-content/themes/'). The version is detected through the 'style.css' file and the version tag.
        
        Besides finding all installed themes, WPScan looks for the 'main theme'. The 'main theme' is the currently active theme that styles the page. To detect this theme, WPScan scans a website's content for links containing the themes folder.
        
        To evade theme detection, we used the same two methods as for the plugins, as described in Section \ref{sec:plugin_detection}. This includes logging fake themes.
        
        Evading the main theme detection would require changing links inside the website, which would lead to a loss of the theme style. Since this is a significant loss of functionality, we decided to not change it.
    
    \subsection{Version Detection}
    \label{sec:version_detection}
        Version detection refers to detecting the running version of the WordPress installation. WP displays the version in different ways. WPScan can detect the following:
        \begin{itemize}
            \item \textit{WordPress Head}: WordPress prints the version in the head of the website.
            \item \textit{WordPress Generator}: WordPress adds the version the the XML generator. 
            \item \textit{Version Query}: WordPress also shows the version as a query argument for core style and script files.
            \item \textit{Fingerprinting}: All core JavaScript, CSS and JSON files of a WordPress version have a unique hash with which the version can be detected.
            \item \textit{File Content}: Both the 'install.php' and 'load\_styles.php' scripts output the version.
        \end{itemize}
    
        To break the version detection, we removed the WP head and generator as well as the version query arguments. We added a space to all the core files to change the hashes, so fingerprinting is no longer possible. Finally, we removed the lines of code in the scripts, thus they no longer display the versions.
        
        Regarding simulation, it should not be possible to manipulate the WordPress head or version query. In theory, it is possible to change the generator response. Fingerprinting could be simulated by adding core files from older versions. This is, however, not desirable since WordPress should be up to date. The php file content of the 'install.php' could be changed instead of removed, but the 'load\_styles.php' appends versions to the links to get the correct file and therefore also could not be simulated.
        
    \subsection{User Enumeration}
    \label{sec:user_enum}
        User enumeration tries to detect existing usernames to log into the WordPress instance. For this WPScan uses different functionality that WordPress offers:
        \begin{itemize}
            \item \textit{REST API}: WordPress allows to query users through the available REST API.
            \item \textit{JSON API}: The JSON API also makes it possible to query users.
            \item \textit{Author Class}: When authors of posts answer comments, they are marked with an author class.
            \item \textit{RSS Author Tag}: The RSS feed contains a specific tag naming the author.
            \item \textit{Login Error}: Login error messages can display if the user exists.
            \item \textit{URL Queries}: WordPress allows to do URL queries with the user id.
            \item \textit{Author URL}: WordPress adds an author link to posts that contains the username.
        \end{itemize}
        Completely evading user enumeration comes with a loss of functionality. The REST and JSON API are disabled, which restricts the ability of applications to interact with your WordPress site. URL queries are forbidden. The author class, author link and RSS author tag are removed, which reduces the visibility of authors. All login errors are disabled, which might make troubleshooting more difficult. This could potentially also effect search engine optimization.
        
        With the exception of the RSS Author Tag, all other areas can not be used for simulation purpose since it would restrict the functionality. It might be possible to create a user purposely as kind of a 'honeypot' user. However, the plugin removes all user enumeration. The RSS Author Tag though could be overwritten with an arbitrary name.

\section{Discussion}

    We tested our plugin in a test environment. Details and the WPScan output can be found in the Git repository (https://github.com/dfki-in-sec/SCANTRAP). The results show that the proposed strategies work for all four scans, with the exception of the detection of the main theme, as described in Section \ref{sec:theme_detection}.
    
    One downside regarding our methodology is that some of the functionality, that might be desired by the users, are restricted with the use of our plugin. Especially user enumeration leads to potentially undesirable loss. However, user enumeration could be seen as more of a privacy concern. Every defender would need to decide for themselves if they put more importance on functionality or privacy. One exception to this assumption is that WordPress usernames are also login names and WordPress does not restrict login tries. This makes user enumeration interesting in the context of login brute force attacks. As mentioned in \ref{sec:theme_detection}, the main theme detection is one example of full loss of the main functionality 'style'. However, it has to be said that themes only make up 6\% of the existing vulnerabilities \cite{WPVulnstats}. A similar problem could occur if plugins require URLs to the plugin folder on any pages. Another problem that could occur is with search engine optimization, since the plugin removes author links and manipulates status codes to certain folders and files. 
   
    We are aware of two limitations in this methodology: the first is our testing environment, as it is not a real WordPress blog. We tested with only a limited amount of plugins, themes, users and content. Due to a large number of existing plugins alone, it would not be feasible to test everything. The second is the restrictions on CMSs and CMS scanners. Our plugin is only focusing on WordPress and WPScan. As shown in Section \ref{sec:background} there are more CMSs and scanners that might require adjustment to our method.
    
    Cyber deception is supposed to deter attackers and make attacking harder. It will not remove vulnerabilities or definitely hinder an attacker from targeting the CMS. Therefore, having other security measures, like up-to-date software, is still required. However, it might discourage attackers completely from seeing the CMS as a target. More likely, it will require more resources for additional manual reconnaissance or lead attackers to the wrong conclusion about existing vulnerabilities. This might be helpful as a defence method for overlooked vulnerabilities. 


\section{Related Work}
\label{sec:related_work}
    While research in regards to securing web content and CMSs has been conducted, we have found little in regards to using cyber deception, specifically for CMSs. 
    
    Jagamogan et al. \cite{9617087} conducted a review on different penetration testing approaches on CMS. It underlines the likelihood of CMSs being targeted. While the focus lies on exploitation frameworks, it is apparent that a lot are based on different scanning methods. Specifically, it is mentioned that WPScan and other CMS scanners were used for attacks. 
  
    The paper by Jerković et al. \cite{7522359} describes in more detail the problems and different vulnerabilities of CMSs. Such as easy CMS identification and a large base of unsecured and out-of-date plugins. For protection, the paper lists different best practices, such as content delivery networks (CDN), regular updating and logging. They conclude that an up-to-date CMS is not sufficient enough to prevent attacks. Their solution for this is hosting in CDN, which requires additional services and potential costs.
  
   Efendi et al. \cite{8819492} highlighted the use of deception techniques in the domain of web applications. Their study describes web application vulnerabilities and attacks and deception techniques. Their focus lies on the detection of different attacks on the application layer. The concentration is on services and processes like SSH, Javascript and MySQL and attacks such as Denial of Service and SQL Injections. In comparison, we are not focusing on detecting specific attacks but rather hindrance of successful reconnaissance and detection of such in the specific domain of CMSs.  
    
    Valenza et al. \cite{inproceedings} developed a prototype that attacks vulnerability scanners. In their attacker model, the attacker becomes the victim and the defender the attacker. The idea is to use vulnerabilities of the vulnerability scanner application to attack it through HTTP responses. In regards to attacks, their focus is on XSS and exploitation to get access to the machine. Our paper is based on the same idea that vulnerability scanners can not be attacked or manipulated. Instead of finding vulnerabilities in scanners, however, our work focuses on finding structures to manipulate the outcome of the scanner's results.
    
    A blog post from Medium Talk \cite{medium_talk} describes two methods to hide the login URL. The first is to change the name of the login file, the second is to use a cookie to control access to the login page. A similar approach to this is taken by the 'Hide My WP Ghost - Security Plugin' \cite{wpghost}. This plugin allows changing and hiding all major paths and files as well as some additional features. 
 
    The blog post from \textit{Cyberpunk} \cite{WPScanEvasion} that we have built our plugin on, is to our knowledge one of a few comprehensive resources on using cyber deception for CMSs. The post focuses specifically and only on WordPress and WPScan. The first part of the blog describes the usage of WPScan. The second part goes into the defence against detection. For each area, it gives a short description of how WPScan finds information and a block of code that should prevent it. Some of the defence methods require changes outside of WordPress. The focus of the blog post lies in hiding information. We found the listings of sources used by WPScan incomplete and therefore, some of the code snippets did not lead to the desired effect on their own. We build upon this by adding defence against additional sources, removing the need to manually change settings in web servers and adding upon the idea by including additional cyber defence techniques like simulation and honeytokens.
    
    Another blog post that deals with WPScan obfuscation is by \textit{shift8} \cite{WPScanEvasion2}. It deals specifically with blocking  WPScan with Nginx. It has similar content and drawbacks to the previously described blog post. However, it additionally mentions the approach to blocking plugin and theme folders by changing response codes to 404 status codes.

    The first posts on blocking web CMS scanners, we found, are about user enumeration \cite{wpbullet} and blocking WPScan \cite{gorgolak}. Both posts can be considered partly outdated and incomplete. Both focus on dissimulation and do not include simulation.

        \begin{small}
        \begin{table*}[t]\centering
            
        \caption{Comparing different WordPress security plugins regarding their dissimulation and simulation features.}
        \label{tab:secplugins} 
        {
        \footnotesize
        \renewcommand{\arraystretch}{2}
            \begin{tabular}{|c|c|c|c|c|c|c|c|c|c|c|c|c|} 
             \hline
             Plugin / Capability &
             \rot{90}{\shortstack{Hide/Fake \\ Users}} & 
             \rot{90}{\shortstack{Hide/Fake \\ Plugins}} & 
             \rot{90}{\shortstack{Hide/Fake \\ Themes}} & 
             \rot{90}{\shortstack{Hide/Fake \\ 'wp-content'}} & 
             \rot{90}{\shortstack{Hide/Fake \\ Login Path}} & 
             \rot{90}{\shortstack{Hide/Fake \\ Login Error}} & 
             \rot{90}{\shortstack{Hide/Fake \\ Version}} & 
             \rot{90}{\shortstack{Hide/Fake \\ Cookies}} & 
             \rot{90}{\shortstack{Hide \\ WordPress}} & 
             \rot{90}{\shortstack{Hide/Fake \\ robots.txt}} &
             \rot{90}{\shortstack{Detect/Block \\ Attackers}} \\
             \hline
             SCANTRAP & \ \cmark/- & \ \cmark/\ \cmark & \ \cmark/\ \cmark & -/- & -/- & \ \cmark/- & \ \cmark/- & -/- & - & -/- & \ \cmark/- \\
             \hline
             \shortstack{Blackhole \\ \cite{blackhole}}  & -/- & -/- & -/- & -/- & -/- & -/- & -/- & -/- & - & -/\ \cmark & \ \cmark/\ \cmark \\
             \hline
             \shortstack{WP Ghost \\ \cite{wpghost}} & \ \cmark/- & \ \cmark/- & \ \cmark/- & \ \cmark/- & \ \cmark/\ \cmark & -/- & -/- & -/- & \ \cmark & -/- & \ \cmark/\ \cmark\\
             \hline
             \shortstack{Don Security \\ \cite{donsec}} & \ \cmark/- & \ \cmark/- & -/- & -/- & -/- & -/- & \ \cmark/- & -/- & - & \ \cmark/- & -/- \\
             \hline
             \shortstack{Stop User Enum \\ \cite{userenum}} & \ \cmark/- & -/- & -/- & -/- & -/- & -/- & -/- & -/- & - & -/- & -/- \\
             \hline
             \shortstack{WP Smart Security\\ \cite{smart}} & -/- & \ \cmark/- & \ \cmark/- & \ \cmark/- & \ \cmark/- & \ \cmark/- & \ \cmark/\ \cmark & -/- & \ \cmark & -/- & \ \cmark/\ \cmark \\
             \hline
             \shortstack{Titan \\ \cite{antispam}} & -/- & -/- & -/- & -/- & \ \cmark/- & -/- & \ \cmark/- & -/- & - & -/- & \ \cmark/\ \cmark \\
             \hline
             \shortstack{tinyShield (deprecated) \\ \cite{tinyshield}} & \ \cmark/- & -/- & -/- & -/- & -/- & -/- & -/- & -/- & - & -/- & \ \cmark/\ \cmark \\
             \hline
             \shortstack{block-wpscan \\ \cite{blockwpscan}} & -/- & -/- & -/- & -/- & -/- & -/- & -/- & -/- & - & -/- & \ \cmark/\ \cmark \\
             \hline
             \shortstack{NovaSense \\ \cite{novasense}} & -/- & -/- & -/- & -/- & -/- & -/- & -/- & -/- & - & -/- & \ \cmark/\ \cmark \\
             \hline
             \shortstack{Astra Security \\ \cite{astra}} & -/- & -/- & -/- & -/- & -/- & -/- & -/- & -/- & - & -/- & \ \cmark/\ \cmark \\
             \hline
             \shortstack{BlogSafe Honeypot \\ \cite{blogsafe}} & -/- & -/- & -/- & -/- & -/- & -/- & -/- & -/- & - & -/- & \ \cmark/- \\
             \hline
            \end{tabular}
        \renewcommand{\arraystretch}{1}
        }
        \end{table*}
        \end{small}
        
    We have found a few plugins in the WordPress store that have similar security and deception ideas. Table \ref{tab:secplugins} gives an overview of the different functions offered by them. We are mainly looking at hiding and faking, so dissimulation and simulation respectively. Additionally, the detection and blocking of attackers is a broad term for whitelisting, blacklisting IPs, as well as preventing brute-force attacks. Some of the plugins might also include some additional functionality.
    
    'Blackhole for Bad Bots' creates honeysites and blocks IP addresses \cite{blackhole}. In a similar manner, 'tinyShield' (deprecated) adds functionality of white- and blacklisting \cite{tinyshield}, so does 'NovaSense' \cite{novasense}. Another plugin called 'BlogSafe Honeypot' tracks failed URL requests and login attempts \cite{blogsafe}. 'Astra Security Suite' contains IP blocking mechanisms, logging, attack prevention and more \cite{astra}. This is similar to the plugin 'WP Smart Security', however, it has not been updated in a while \cite{smart}. The 'Titan Anti-spam \& Security' plugin does spam detection, has a login rate limiter and uses WPScan \cite{antispam}. The WPScan database is also used by 'Jetpack Protect' \cite{jetpack}. Now, all of these plugins have their own technique for securing WordPress with a strong focus on attack prevention and less on scanner deception. 
    
    We found three plugins that focus more on scanner evasion with similar methods to our plugin. 'Stop User Enumeration' focuses only on user enumeration \cite{userenum}. The plugin 'block-wpscan' is based on detecting browser languages and user agents, however, it is no longer maintained and WPScan can evade this type of detection \cite{blockwpscan}. Finally, the plugin 'Don Security' \cite{donsec} seems to be closest to our plugin. Its goal is to prevent user enumeration, version detection and plugin detection, specifically for WPScan. It seems to be based on some of the mentioned outdated blog posts and is also no longer maintained.

\section{\uppercase{Conclusions}}
\label{sec:conclusion}
    In this paper, we have presented a WordPress plugin that manipulates the output of WPScan scans, in order to counter reconnaissance. This plugin can hide the existence of plugins, themes and users from the enumeration process and present false plugins and themes. The results support the idea that cyber deception and obfuscation methods are applicable to the CMS enumeration process, and can also help to detect attacks at an early stage. This could be effective to protect from 'script kiddies', bots or attackers that use automation to gather information. Future work could generalize this approach and transfer it to other CMS than WordPress. It should be noted that scanning tools could be adapted to be more resilient against cyber deception, representing a cat-and-mouse game. 

\section*{\uppercase{Acknowledgements}}
This research was supported by the German Federal Ministry of Education and Research (BMBF) through the Open6GHub project (Grant 16KISK003K).

\bibliographystyle{apalike}
{\small
\bibliography{literature}}

\end{document}